\documentclass[10pt, doublecolumn]{IEEEtran}
\usepackage{graphicx}
\usepackage{caption}
\usepackage[ruled]{algorithm2e}
\ifCLASSOPTIONcompsoc
  \usepackage[caption=false,font=normalsize,labelfont=sf,textfont=sf]{subfig}
\else
  \usepackage[caption=false,font=footnotesize]{subfig}
\fi
\usepackage{indentfirst}

\usepackage{amsmath}
\allowdisplaybreaks[4]
\usepackage{amssymb}
\usepackage{times}
\usepackage{mathtools}
\usepackage{psfrag}
\usepackage{cite}
\usepackage{lastpage}
\usepackage{fancyhdr}
\usepackage{color}
 \usepackage{amsthm}
\usepackage{bigints}
\newtheorem{Theorem}{Theorem}
\newtheorem{Remark}{Remark}

\newcommand{\Equ}[1]{
  \begin{align}
    #1
  \end{align}}

\begin{document}
\title{Stochastic Geometry Based Modeling and Analysis on Network NOMA in Downlink CoMP Systems}
\author{Yanshi Sun, \IEEEmembership{Member, IEEE}, Zhiguo Ding, \IEEEmembership{Fellow, IEEE}, Xuchu Dai, Momiao Zhou, \IEEEmembership{Member, IEEE} , Zhizhong Ding
\thanks{
Y. Sun, M. Zhou and Z. Ding are  with the School of Computer Science and Information
Engineering, Hefei University of Technology, Hefei, 230009, China. (email: \{sys, mmzhou, zzding\}@hfut.edu.cn).

Z. Ding is with Department of Electrical Engineering and Computer
Science, Khalifa University, Abu Dhabi, UAE, and Department of Electrical
and Electronic Engineering, University of Manchester, Manchester, UK.
(email: zhiguo.ding@manchester.ac.uk).

X. Dai is with the CAS Key Laboratory of Wireless-Optical Communications, University of Science and Technology of China, Hefei, 230026, China. (email: daixc@ustc.edu.cn).
}
\vspace{-2em}}

\maketitle
\begin{abstract}
This paper investigates the performance of network non-orthogonal multiple access (N-NOMA) in a downlink coordinated multi-point (CoMP) system. In the  considered N-NOMA scheme, multiple base stations (BSs) cooperatively serve a CoMP user, meanwhile, each BS serves additional NOMA users by occupying the same resource block allocated to the CoMP user. The locations of the BSs and users are modeled by stochastic geometric models and the interference from the whole network is considered. Through rigorous derivations, the outage probabilities achieved by the CoMP and NOMA users are obtained, respectively. Numerical results are provided to verify the accuracy of the analytical results and also demonstrate the superior performance of  N-NOMA compared to orthogonal multiple access (OMA) based CoMP scheme.
\end{abstract}
\begin{IEEEkeywords}
Network NOMA (N-NOMA), coordinated multi-point (CoMP), stochastic geometry, outage probability.
\end{IEEEkeywords}
\section{Introduction}
\IEEEPARstart{I}n wireless communications, cell-edge users' data rates are more difficult to guarantee compared to cell-center users, because cell-edge users usually suffer from more severe path losses  and interferences. Coordinated multi-point (CoMP) technique, which utilizes the cooperation among spatially distributed base stations (BSs), can improve the performance of the cell-edge users\cite{irmer2011coordinated,maccartney2019millimeter}. However, conventional CoMP is based on orthogonal multiple access (OMA), which results in low spectral efficiency. For example, when multiple BSs cooperatively serve a user, each BS has to allocate a channel resource block (RB) to this user and prohibit other users from accessing this RB.

To address this drawback, network NOMA (N-NOMA) has been proposed for CoMP systems and has attracted significant research attentions\cite{choi2014non,ali2018downlink,sys2017nnomafeasibility}. The key idea of N-NOMA is that when multiple BSs cooperatively serve a cell-edge user (namely CoMP user), each BS serves additional cell-center users (namely NOMA users) simultaneously by occupying the same resource block allocated to the CoMP user. It has been shown that N-NOMA can significantly improve the connectivity and spectral efficiency of CoMP systems, compared to OMA based schemes.

{\color{black}However, most of the papers on N-NOMA focused on the scenarios where the number {\color{black}and/or} locations of communication nodes are given and the interference from whole network are not taken into consideration\cite{
elhattab2022joint,elhattab2022ris,wang2020power,elhattab2020joint}}. Thus, the impact of the spatial randomness and interference on performance of N-NOMA were not well revealed. To address this issue, sporadic efforts have been made by applying tools from stochastic geometry\cite{haenggi2012stochastic}. In \cite{sys2019PCP}, Poisson cluster point process (PCP) was applied to model and analyze the performance of uplink N-NOMA. In \cite{zhang2020performance}, a similar model was applied to evaluate the performance of downlink N-NOMA in a mmWave system. The application of N-NOMA to vehicular networks was studied in \cite{sun2021outage}, where Poisson line Cox process (PCL) is applied.
{\color{black}The performance of downlink N-NOMA in two-tier heterogeneous cloud radio access networks (H-CRANs) was investigated in \cite{elhattab2020comp},
by modeling the locations of remote radio heads (RRHs) as homogeneous Poisson point processes (PPPs).
Note that in \cite{zhang2020performance, sun2021outage,elhattab2020comp}, only two cooperating BSs/RRHs are considered, which lacks of generality.}

{\color{black}To fulfill the above drawback, this paper studies the performance of a more general downlink N-NOMA scenario, {\color{black}where the non-coherent joint transmission (NC-JT) technique \cite{tanbourgi2014tractable} is adopted}. The contributions of this paper are listed as follows.
\begin{itemize}
  \item A PCP model is applied to model the locations of BSs and users, where arbitrary number of cooperating BSs is considered, which significantly generalize the considered scenario of the existing work on stochastic geometric modeling and analysis of N-NOMA \cite{zhang2020performance, sun2021outage,elhattab2020comp}.
  \item Note that performance analysis of this paper is more challenging than the existing work, because it is necessary to jointly characterize the randomness of small and large scale fadings for multiple
      cooperating channel links, whose distribution is much more complex than existing work.  By giving the distribution of the sum of cooperating BSs' channel gains in the form which is beneficial for the application of theories from stochastic geometry, the expressions for the outage probabilities of CoMP and NOMA users are obtained, which contributes to the existing literature.
  \item The accuracy of the developed analytical results can be strictly guaranteed by properly choosing three approximation parameters, termed ``$N$", ``$M_A$" and ``$K_A$". With the developed analysis, system designers can conveniently obtain the trend of how different system parameters impact performance, getting rid of cumbersome computer simulations.
\end{itemize}}

\section{System model}
Consider a downlink N-NOMA scenario as shown in Fig. \ref{system_model},  the locations of the BSs are modeled as a homogeneous PPP with intensity $\lambda_c$, denoted by $\Phi_c=\{x_i\}$, where $x_i$ is the location of the $i$-th BS. This paper focuses on two types of users, namely NOMA users and CoMP users.
The NOMA users are near to the BSs, and can be modeled as a PCP, where $\Phi_c$ acts as the parent process\cite{haenggi2012stochastic}. More specifically, in the $i$-th cluster, there are $K$ NOMA users randomly and uniformly distributed in the disc centered at $x_i$ with radius $\mathcal{R}_c$, and the $k$-th NOMA user is denoted by $U_{i,k}$, $\{1,\cdots,K\}$.
The coordinate of $U_{i,k}$ is given by $x_i+y_{i,k}$\footnote{{\color{black}Please note that $x_i$ and $y_{i,k}$ denote two dimensional coordinates.}}.
The CoMP users are defined as the users whose distances from all BSs are larger than a predefined $\bar{\mathcal{R}}$. This paper consider a user centric N-NOMA scheme,  where each CoMP user invites the BSs whose distances are not larger than $\mathcal{R}_{\mathcal{D}}$ to cooperatively  serve it. Without loss of generality, the following of this paper will focus on a typical CoMP user denoted by $U_0$ whose location is set at the origin. And hence the cooperating BSs are located in the circular ring denoted by $\mathcal{C}$ which is centered at the origin, where the outer disc is denoted by $\mathcal{D}$ with radius $\mathcal{R}_{\mathcal{D}}$, and the inner disc is denoted by $\mathcal{D}'$ with radius $\bar{\mathcal{R}}$, as shown in Fig. \ref{system_model}. In addition, it is noteworthy that assuming the nearest BS to the CoMP user is further than $\bar{\mathcal{R}}$ is equivalent to conditioning on that there is no point of $\Phi_c$ drops in $\mathcal{D}'$.

In addition to serving the CoMP user, by using the same resource block, each cooperative BS simlutaneously serves a NOMA user with largest channel gain from its cluster, denoted by:
\Equ{
   k^*_i= \arg\min\limits_{k}\frac{|h_{i,U_{i,k}}|^2}{L(||y_{i,k}||)},
}
 where $h_{i,U_{i,k}}$ denotes the small scale Rayleigh fading from the $i$-th BS to $U_{i,k}$, and $L(||y_{i,k}||)$ is the large scale fading. Particularly, $\frac{\eta}{L(||y_{i,k}||)}$ is used as the  path loss model in this paper, where $L(||y_{i,k}||)=||y_{i,k}||^{\alpha}$ and $\eta=\frac{c^2}{16\pi^2f_c^2}$ is the coefficient which is relevant to the carrier frequency $f_c$ ($c$ is the speed of light ), $\alpha$ is the large scale path loss exponent.
\begin{figure}[!t]
\vspace{-1em}
\setlength{\abovecaptionskip}{0em}   %
\setlength{\belowcaptionskip}{-1em}   %
  \centering
  \includegraphics[width=2.7in]{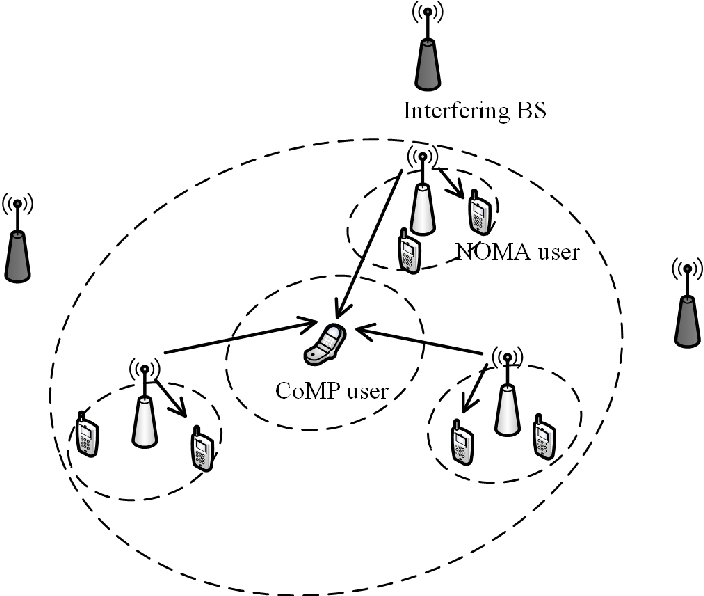}\\
  \caption{Illustration of the system model. $K=2$}\label{system_model}
  \label{system_model}
\end{figure}
{\color{black}The transmitted signal {\color{black}at the $m$-th subcarrier}{\footnote{{\color{black}It is assumed that orthogonal frequency division multiplexing (OFDM) waveform with $N_c$ carriers is applied in the considered transmission.}}} by BS $i$ in disc $\mathcal{D}$ is given by:
\begin{align}
  x_i[m]=\beta_0\sqrt{P_s}s_0[m]+\beta_1\sqrt{P_s}s_i[m],
\end{align}
where superposition coding (SC) is applied, $s_0[m]$ is the signal intended for the CoMP user $U_0$ at the $m$-th subcarrier, $s_i[m]$ is the signal intended for the NOMA user of BS $i$ at the $m$-th subcarrier. $s_0[m]$ and $s_i[m]$ are independent with each other, and the signal powers of $s_0[m]$ and $s_i[m]$ are normalized. $P_s$ is the transmission power at each subcarrier (even power allocation for different subcarriers is assumed), $\beta_0$ and $\beta_1$ are the power allocation coefficients,
and $\beta_0^2+\beta_1^2=1$.

At the receiver, {\color{black}the observed signal at the $m$-th subcarrier} by the CoMP user is given by:
\begin{align}\label{receive_0}
  y_0[m]=\!\!\!\!\!\!&\sum_{x_i\in \Phi _c\cap \mathcal{D}}{\!\!\!\!\frac{g_i[m]e^{j2\pi m \frac{\nu_i}{N_c}}}{\sqrt{L\left( ||x_i|| \right)}}\left( \beta _0\sqrt{P_s}s_0[m]\!+\!\beta _1\sqrt{P_s}s_i[m] \right)}\\\notag
  &+\!\!\!\sum_{x_i\in \Phi _c\backslash \Phi _c\cap \mathcal{D}}{\frac{g_i[m]e^{j2\pi m \frac{\nu_i}{N_c}}}{\sqrt{L\left( ||x_i|| \right)}}\sqrt{P_s}\tilde{s}_i[m]}+n_0[m],
\end{align}
where $\tilde{s}_i[m]$ is the interfering signal at the $m$-th subcarrier from the BS which is outside disc $\mathcal{D}$, whose power is normalized, i.e., $\mathbb{E}\{|\tilde{s}_i[m]|^2\}=1$;
$n_0[m]$ is the gaussian noise, $n_0[m] \sim \mathcal{CN}(0,\sigma^2)$, $\sigma^2$ is the noise power;
$g_i[m]$ is the small scale Rayleigh fading from BS $i$  to the CoMP user. It is assumed that $g_i[m]$ remains constant, i.e., $g_i[m]=g_i$ within the coherence bandwidth\cite{tanbourgi2014tractable} (usually a few tens times the subcarrier spacing).
In this paper, NC-JT is considered, where the cooperating BSs jointly transmit the same signal to the CoMP user, without prior phase mismatch correction and tight synchronization\cite{tanbourgi2014tractable}{\footnote{{\color{black}This paper focuses on the scenario where the cooperating BSs are connected to a central processing unit (CPU). The CPU distributes the signal intended for the CoMP user to the cooperating BSs, thus the cooperating BSs don't need to exchange the data of the CoMP user.}}}.
As a result, there is a time offset $\nu_i$ in the time domain for each channel link, corresponding to a phase shift in the frequency domain as shown in (\ref{receive_0}).
Note that, $g_i$s from the cooperating BSs are known at the CoMP user, whereas the information of $\nu_i$s are not available. It is also assumed that $\nu_i$s are independent across different BSs.
Interestingly, by applying NC-JT, the effect of $\nu_i$s can be removed, and a received power boost can be obtained, which is known as the cyclic delay diversity (CDD). For more details on how the CDD is obtained, interested readers can refer to Appendix A in \cite{tanbourgi2014tractable}. By following the similar steps as  in \cite{tanbourgi2014tractable}, the signal to interference plus noise ratio (SINR) of the CoMP user can be expressed by:
\begin{align}
  \text{SINR}_0\!=\!\frac{\beta^2_0\sum\limits_{x_i\in \Phi _c\cap \mathcal{D}}\frac{||g_i||^2}{L(||x_i||)}}
                {\beta_1^2\sum\limits_{x_i\in \Phi _c\cap \mathcal{D}}\frac{||g_i||^2}{L(||x_i||)}\!+\!
                \sum\limits_{x_i\in \Phi _c\backslash \Phi _c\cap \mathcal{D}}\frac{||g_i||^2}{L(||x_i||}
                \!+\!\frac{1}{\rho}}
\end{align}
where the NOMA users' signals are treated as interferences, and $\rho=\frac{\eta P_s}{\sigma^2}$.}

It is assumed that the CoMP user has the knowledge of the channel state information (CSI) of the cooperating BSs. However, in practice, it is reasonable to assume that each NOMA user only has the CSI of its serving BS, for the sake of reducing system overhead. For the NOMA user served by BS $i$, i.e., $U_{i,k_i^*}$, it first decodes the CoMP user's signal with  the following SINR:
\begin{align}\label{SINRi0}
  \text{SINR}_{i,0}\!=\!\frac{\frac{|h_{i,U_{i,k_i^*}}|^2}{L(|||y_{i,k_i^*}|^2)}\beta_0^2}
                         {\frac{|h_{i,U_{i,k_i^*}}|^2}{L(|||y_{i,k_i^*}|^2)}\beta_1^2+\!\!\!\!\!\!\!
                         \sum\limits_{x_j \in \Phi_c\backslash x_i}\frac{|h_{j,U_{i,k_i^*}}|^2}{L(||y_{i,k_i^*}+x_i-x_j||^2)}\!+\!\frac{1}{\rho}}.
\end{align}
If successful, the NOMA user will remove the CoMP user's signal and then decode its own signal with the following SINR:
\begin{align}\label{SINRiii}
  \text{SINR}_{i,i}=\frac{\frac{|h_{i,U_{i,k_i^*}}|^2}{L(|||y_{i,k_i^*}|^2)}\beta_1^2}
  {
  \sum\limits_{x_j \in \Phi_c\backslash x_i}\frac{|h_{j,U_{i,k_i^*}}|^2}{L(||y_{i,k_i^*}+x_i-x_j||^2)}+\frac{1}{\rho}}.
\end{align}

{\color{black}Note that, for tractable analysis and focusing on characterizing the effect of the random topology of BSs, this paper makes the assumption that all the communication nodes are equipped with a single antenna. This assumption is applicable in many scenarios, such as Internet-of-things (IoTs), C-RANs and small cells, where BSs/APs in such scenarios are usually limited in cost and size. Besides, considering the scenarios with multiple antennas will lead to other research challenges, such as how to group
users or how to design beamformers, which are beyond the scope of this paper\cite{sys2022QD}.}
\section{Performance Analysis}
\subsection{CoMP User's Outage Probability}
The outage probability achieved by the CoMP user is given by:
\begin{align}
  P_{0}^{out}=\text{Pr}\left( \text{SINR}_0 <\epsilon_0\right),
\end{align}
where $\epsilon_0=2^{R_0}-1$, $R_0$ is the target rate of the CoMP user.

To evaluate $P_{0}^{out}$, it is necessary to characterize the distribution of the channel gains of the cooperating BSs. For a randomly chosen cooperating BS $i$, let $\tilde{g_i}=\frac{|g_i|^2}{L(||x_i||)}$ , where $x_i$ is uniformly distributed in the circular ring $\mathcal{C}$. By applying Gaussian-Chebyshev approximation \cite{hildebrand1987introduction}, the {\color{black}probability density function (pdf)} of $\tilde{g}_i$ can be approximated as a sum of exponentials:
\begin{align}\label{sum_exp}
  f_{\tilde{g}_i}(z) \approx  \sum^{N}_{n=1}w_nd_n  e^{-d_nz},
\end{align}
where
\begin{align}
   &w_n=\frac{N}{2\pi}\sqrt{1-\theta_n^2}\left(\frac{\mathcal{R}_{\mathcal{D}}-
   \bar{\mathcal{R}}}{\mathcal{R}_{\mathcal{D}}+\bar{\mathcal{R}}}\theta_n+1\right),\\
   &d_n=\left(\frac{\mathcal{R}_{\mathcal{D}}-\bar{\mathcal{R}}}{2}\theta_n+\frac{\mathcal{R}_{\mathcal{D}}+\bar{\mathcal{R}}}{2}\right)^{\alpha},\\
   &\theta_n=\cos\left(\frac{2n-1}{2N}\pi\right),
\end{align}
and $N$ is Gaussian-Cheyshev parameter.

{\color{black}The proof for (\ref{sum_exp}) is similar to the proof for Lemma 1 in \cite{sys2019PCP}, and is omitted in this paper due to space limitations. As shown in \cite{sys2019PCP}, the approximation error in  (\ref{sum_exp}) decreases rapidly with $N$, and hence the accuracy of the approximation can be ensured by a properly chosen $N$.}

According to the conditional property for PPP, if it is assumed that there are $M$ BSs in the circular ring $\mathcal{C}$, then it can can be concluded that the $M$ BSs are independently and uniformly distributed in the ring \cite{haenggi2012stochastic}. The sum of the channel gains of these BSs is denoted by
$G_M=\sum_{i=1}^M\tilde{g}_i$. By using the same method as in \cite{ding2018coexistence}, the  pdf of $G_M$ can be expressed as follows:
\begin{align}\label{G_M_pdf}
  f_{G_M}(x)\approx&\sum_{\substack{k_1+\cdots+k_N=M}}{M \choose k_1, \cdots, k_N}
  \times\\\notag
  &\sum\limits_{n=1,k_n\neq0}^{N}\sum\limits_{i_n=0}^{k_n-1}\frac{
  A_{i_n}x^{k_n-i_n-1}e^{-d_nx}}{(k_n-i_n-1)!},
\end{align}
where
\Equ{A_{i_n}=\frac{1}{i_n!}\frac{d^{i_n}}{ds^{i_n}}\left[ (s+d_n)^{k_n}q(s) \right] |_{s=-d_n},}
and
\Equ{
  q(s)=\prod_{n=1,k_n\neq0}^{N}\frac{w_n^{k_n}d_n^{k_n}}{(d_n+s)^{k_n}}.}

Based on the above results, the outage achieved by the CoMP user can be obtained as
shown in the following theorem.
\begin{Theorem}
By applying NC-JT in N-NOMA, outage probability achieved by the CoMP user can be approximated as:
\begin{align}
	&P_{0}^{out} \approx \sum_{m=0}^{M_A}{\frac{(\lambda _cS_{\mathcal{C}})^m}{m!}}e^{-\lambda _cS_{\mathcal{C}}}\!\!\!\!\!\!\!\sum_{k_1+\cdots +k_N=m}{
	m \choose {k_1,\cdots ,k_N}}\\\notag
	&\times \sum_{n=1,k_n\ne 0}^N\sum_{i_n=0}^{k_n-1}{\frac{A_{i_n}}{d_{n}^{k_n-i_n}}\sum_{k=0}^{K_A}{\frac{\left( -\mu \right) ^{k_n-i_n+k}\mathcal{L} ^{\left( k_n-i_n+k \right)}\left( \mu \right)}{\Gamma \left( k_n-i_n+k+1 \right)}}} ,
\end{align}
where
${S_{\mathcal{C}}=\pi(\mathcal{R}_{\mathcal{D}}^2-\bar{\mathcal{R}}^2)}$,
${\mu =\frac{d_n\epsilon _0}{\beta _{0}^{2}-\beta _{1}^{2}\epsilon _0}}$,
and $\mathcal{L}^{(n)}(\mu)$ is the $n$-th derivative of $\mathcal{L}(\mu)$, which is given by:
\Equ{\mathcal{L} (\mu)\!=\exp \bigg(\!\!-\frac{\mu}{\rho} \!-\!\frac{2\pi \lambda \mu \mathcal{R} _{\mathcal{D}}^{2-\alpha}}{\alpha -2}\,_2F_1\left(\! 1\!-\!\frac{2}{\alpha},1;2\!-\!\frac{2}{\alpha};\!\frac{-\mu}{\mathcal{R} _{\mathcal{D}}^{\alpha}} \right)\!\bigg),}
$_2F_1(\cdot)$ is the gaussian hypergeometric function, $\Gamma(n)=(n-1)!$, and $M_A$ and $K_A$ are the parameters which controls the accuracy of the approximation.
\end{Theorem}
\begin{IEEEproof}
Let $M$ be the number of cooperating BSs in circular ring $\mathcal{C}$, without loss of generality, the indexes of these BSs are set to be from $1$ to $M$.  Note that, $M$ is a random variable and the outage probability of CoMP user can be written as:
\begin{align}
 P_{0}^{out}=\sum_{m=0}^{\infty}\text{Pr}(M=m)P^{out}_{0,m},
\end{align}
where
\Equ{
    P_{0,m}^{out}= \text{Pr}\left(
    \frac{\beta^2_0\sum\limits_{i=1}^m\frac{||g_i||^2}{L(||x_i||)}}
                {\beta_1^2\sum\limits_{i=1}^m\frac{||g_i||^2}{L(||x_i||)}+
                \sum\limits_{x_i\in \Phi _c\backslash \Phi _c\cap \mathcal{D}}\frac{||g_i||^2}{L(||x_i||}
                +\frac{1}{\rho}}<\epsilon_0
    \right).
}
$\text{Pr}(M=m)$ can be easily obtained by using the definition of PPP as expressed by:
\begin{align}\label{P_Mm}
\text{Pr}(M=m)=\frac{(\lambda_cS_{\mathcal{C}})^m}{m!}e^{-\lambda_cS_{\mathcal{C}}},
\end{align}
where $S_{\mathcal{C}}$ is the area of $\mathcal{C}$ \cite{haenggi2012stochastic}.

The remaining task is to calculate $P_{0,m}^{out}$. Note that it can be written as:
\Equ{
     P_{0,m}^{out}&=\text{Pr}\left( \sum_{i=1}^m{\frac{|g_i|^2}{L\left( ||x_i|| \right)}}< \frac{\epsilon_0(I_{out}+1/\rho)}{\beta_0^2-\beta_1^2\epsilon_0} \right)\\\notag
     &=\text{Pr}\left( G_M< \frac{\epsilon_0(I_{out}+1/\rho)}{\beta_0^2-\beta_1^2\epsilon_0} \right),
}
where
\Equ{
 I_{out}=\sum_{x\in \Phi _c\backslash \Phi _c\cap \mathcal{D}}{\frac{|g_i|^2}{L\left( ||x_i|| \right)}}.
}
By integrating the pdf of $G_M$ as shown in (\ref{G_M_pdf}) and applying the incomplete Gamma function, the cdf of $G_M$ can be obtained as:
\begin{align}\label{F_G_M}
F_{G_M}(x)\approx&\sum_{\substack{k_1+\cdots+k_N=M}}{M \choose k_1, \cdots, k_N}
\\\notag
&\times\sum\limits_{n=1,k_n\neq0}^{N}\sum\limits_{i_n=0}^{k_n-1}\frac{
A_{i_n}\gamma(k_n-i_n,d_nx)}{(k_n-i_n-1)!d_n^{k_n-i_n}}.
\end{align}
where, $\gamma(s,x)=\int_{0}^xt^{s-1}e^{-t}\,dt$ is the lower incomplete Gamma function.

By applying (\ref{F_G_M}), $P_{0,m}^{out}$ can be further expressed as:
\begin{align}\label{P_G_m}
P_{0,m}^{out} =&\mathbb{E}_{I_{out}}\left\{F_{G_M}\left(\frac{\epsilon_0(I_{out}+1/\rho)}{\beta_0^2-\beta_1^2\epsilon_0}\right)\right\}\\\notag
\approx&\sum_{\substack{k_1+\cdots+k_N=m}}{m \choose k_1, \cdots, k_N}
\sum\limits_{n=1,k_n\neq0}^{N}\sum\limits_{i_n=0}^{k_n-1}\\\notag
&\frac{
A_{i_n}\mathbb{E}_{I_{out}}\left\{\gamma\left(k_n-i_n,d_n\frac{\epsilon_0(I_{out}+1/\rho)}{\beta_0^2-\beta_1^2\epsilon_0}\right)\right\}}{(k_n-i_n-1)!d_n^{k_n-i_n}}.
\end{align}

The next task is to take the expectation over $I_{out}$. Note that $I_{out}$ contains two kinds of randomness, one is the Rayleigh small scale fading, and the other is the random locations of the interfering BSs. The integral form of $\gamma$ makes it challenging to evaluate $\mathbb{E}_{I_{out}}\left\{\gamma\left(k_n-i_n,d_n\frac{\epsilon_0(I_{out}+1/\rho)}{\beta_0^2-\beta_1^2\epsilon_0}\right)\right\}$.
Thus, it is necessary to transform the gamma function into another form which is favorable for the application of the conclusions in stochastic geometry, as follows:
\begin{align}\label{gamma_sum}
 \gamma(s,x)=\Gamma(s)\sum_{k=0}^{\infty}\frac{x^{s+k}e^{-x}}{\Gamma(s+k+1)}.
\end{align}

Based on (\ref{gamma_sum}), by using the same method to apply {\color{black}the probability generating functional (pgfl)} of PPP as in Theorem $1$ in \cite{sysmmwave2018}, the expression for $P_{0,m}^{out}$ can be obtained and the proof is complete.
\end{IEEEproof}
\begin{Remark}
If let
  \Equ{
     \tau(\mu)=-\mu /\rho -\frac{2\pi \lambda \mu \mathcal{R} _{\mathcal{D}}^{2-\alpha}}{\alpha -2}  \,\,_2F_1\left( 1-\frac{2}{\alpha},1;2-\frac{2}{\alpha};\frac{-\mu}{\mathcal{R} _{\mathcal{D}}^{\alpha}} \right),
  }
then $\mathcal{L}(\mu)$ can be expressed as $\mathcal{L}(\mu)=\exp(\tau(\mu))$, and $\mathcal{L}^{(n)}(\mu)$ can be calculated iteratively as follows:
\Equ{
  \mathcal {L}^{(i)}(\mu)=\sum _{j=0}^{i-1}{\binom{i-1 }{ j}}\tau ^{(i-j)}(\mu)\mathcal {L}^{(j)}(\mu).
}
which is helpful to reduce the computational complexity.
\end{Remark}
\subsection{NOMA User's Outage Probability}
The outage probability achieved by the NOMA user served by BS $i${\footnote{BS $i$ is randomly chosen from the cooperating BSs.}} given by:
\begin{align}\label{Def_P_i}
  P_{i}^{out}=1-\text{Pr}\left( \mathrm{SINR}_{i,0}>\epsilon _0,\mathrm{SINR}_{i,i}>\epsilon _i \right),
\end{align}
where $\epsilon_i=2^{R_i}-1$, $R_i$ is the target rate of the NOMA user served by BS $i$.

The following theorem provides the expression for $P_{i}^{out}$.
\begin{Theorem}
The outage probability achieved by user $U_{i,k_i^*}$ can be expressed as:
\begin{align}\label{NOMA_outage_Pro}
&P_{i}^{out}\approx 1\!+\!\!\!\!\!\!\sum_{
	\overset{k_0+\cdots +k_N=K}{
	k_0\ne K}}{
	K \choose k_0,\cdots ,k_N}\bigg(\prod_{n=0,k_n\ne 0}^N\!\!{\tilde{w}_{n}^{k_n}}e^{-k_n\tilde{c}_n\bar{\epsilon}_i\frac{1}{\rho}}\bigg)\\\notag
&\times \exp \left( -2\pi \lambda _c\frac{(\xi\bar{\epsilon}_i )^{\frac{2}{\alpha}}}{\alpha}\mathrm{B}\left( \frac{2}{\alpha},\frac{\alpha -2}{\alpha} \right) \right)\int_{\bar{\mathcal{R}}}^{R_{\mathcal{D}}}
\frac{2d}{\mathcal{R}_{\mathcal{D}}^2-\bar{\mathcal{R}}^2}\\\notag
&\quad\exp\left(2\lambda_c\int_{d-\bar{\mathcal{R}}}^{d+\bar{\mathcal{R}}}
\frac{r\xi\bar{\epsilon}_i}{r^\alpha+\xi\bar{\epsilon}_i}\arccos\left(\frac{r^2+d^2-\bar{\mathcal{R}}}{2rd}\right)\,dr \right)\,dd.
\end{align}
where $\bar{\epsilon}_i=\max \left\{ \frac{\epsilon_0}{\beta _{0}^{2}-\beta _{1}^{2}\epsilon _0},\epsilon_i/\beta _{1}^{2} \right\}$, $N$ is Gaussian-Chebyshev parameter, $\tilde{w}_n {\color{black}=}-\frac{\pi}{2N}\sqrt{1-\theta_n^2}(\theta_n+1)$, $1\leq n \leq N$, $\tilde{w}_0 {\color{black}=}1$,
$\tilde{c}_n {\color{black}=}\left(\frac{\mathcal{R}_c}{2}\theta_n+\frac{\mathcal{R}_c}{2}\right)^\alpha$, $1\leq n \leq N$, $\tilde{c}_0 {\color{black}=}0$,
$k_n \geq0$, $0\leq n \leq N$,
$\xi {\color{black}=} \sum_{n=0,k_n\neq 0}^{N}k_n\tilde{c}_n$,
${K \choose k_0,\cdots,k_N} {\color{black}=}\frac{K!}{k_0!\cdots k_N!}$,
and $B(\cdot)$ is the Beta function.
\end{Theorem}
\begin{IEEEproof}
Note that, there are $K$ users randomly distributed in the disc centered at BS $i$, define:
$
 z_{i,k}=\frac{|h_{i,U_{i,k}}|^2}{L(||y_{i,k}||)},
$
as the unordered channel gain, then we have $z_{i,k_i^*}=\max\{z_{i,1},\cdots, z_{i,K}\}$. Similar to (\ref{sum_exp}), the CDF of the unordered channel gain $z_{i,k}$ can be expressed as:
\begin{align}\label{F_z_ik}
 F_{z_{i,k}}(z)\approx\sum_{n=1}^Nw_ne^{-c_nz},
\end{align}
where $w_n=\frac{\pi}{2N}\sqrt{1-\theta_n^2}(\theta_n+1)$, $c_n=\left(\frac{\mathcal{R}_c}{2}\theta_n+\frac{\mathcal{R}_c}{2}\right)^{\alpha}$, and $\theta_n=\cos\left(\frac{2n-1}{2N}\pi\right)$.

By taking (\ref{SINRi0}) and (\ref{SINRiii}) into (\ref{Def_P_i}), $P_{i}^{out}$ can be expressed as:
\Equ{
    P_{i}^{out}
    &=\text{Pr}\left(z_{i,k_i^*}<\bar{\epsilon}_i(I_{\text{inter}}^D+1/\rho)\right)\\\notag
    &=\mathbb{E}_{I_{\text{inter}}^D}
    \left\{\left(\text{Pr}\left(z_{i,k}<\bar{\epsilon}_i(I_{\text{inter}}^D+1/\rho)\right)\right)^K \left| \Phi_c\cap\right.\mathcal{D}'=\emptyset
\right\},
}
where $\emptyset$ represents the empty set, and
\Equ{
    I_{\text{inter}}^D=\sum\limits_{x_j \in \Phi_c\backslash x_i}\frac{|h_{j,U_{i,k_i^*}}|^2}{L(||y_{i,k_i^*}+x_i-x_j||^2)}.
}

By applying (\ref{F_z_ik}), $P_{i}^{out}$ can be further approximated as
\Equ{
    P_{i}^{out}
         &{\approx} {\mathbb{E}}_{I_{\text{inter}}^D }\left\{ \left(1-\sum^{N}_{n=1}w_n e^{-c_n\bar{\epsilon}_i\left(I_{\text{inter}}^D
         +\frac{1}{\rho}\right)}\right)^K \right\}\\\notag
         &= 1+\sum_{\substack{k_0+\cdots+k_N=K \\ k_0\neq K}}{K \choose k_0, \cdots, k_N} \bigg(\prod_{n=0,k_n\neq 0}^{N} \\\notag
         &\quad\quad\tilde{w}_n^{k_n} e^{-k_n\tilde{c}_n\bar{\epsilon}_i
    \frac{1}{\rho}}\bigg)\times{\mathbb{E}}_{I_{\text{inter}}^D } \left\{e^{-\xi\bar{\epsilon}_i  I_{\text{inter}}^D}\left| \Phi_c\cap\right.\mathcal{D}'=\emptyset\right\}.
}

The last term in the above equation is the Laplace transform of the interference, and can be evaluated as follows:
\begin{align}
&\quad{\mathbb{E}}_{I_{\text{inter}}^D } \left\{e^{-\xi\bar{\epsilon}_i  I_{\text{inter}}^D}\left| \Phi_c\cap\right.\mathcal{D}'=\emptyset\right\}\\\notag
&=\mathbb{E}_{I_{\text{inter}}^D}\left\{\exp\left(\!-\!\!\sum_{x_j\in \Phi_c\backslash x_i}\!\!\!\frac{\xi\bar{\epsilon}_i||h_{j,U_{i,k_i^*}}||^2}{||y_{i,k_i^*}\!+\!x_i\!-\!x_j||^{\alpha}}\right)\bigg|\Phi_c\cap\mathcal{D}'=\emptyset\right\}\\\notag
&\overset{(a)}{=}\mathbb{E}_{\Phi_c}\left\{
\prod_{x_j\in \Phi_c\backslash x_i}\frac{||y_{i,k_i^*}+x_i-x_j||^{\alpha}}{||y_{i,k_i^*}+x_i-x_j||^{\alpha}+\xi\bar{\epsilon}_i}
\bigg|\Phi_c\cap\mathcal{D}'=\emptyset\right\}\\\notag
&\overset{(b)}{=}\mathbb{E}_{x_i,y_{i,k_i^*}}\!\left\{\exp\left\{\!-\lambda_c\!\int_{||x||>\bar{\mathcal{R}}}
\frac{\xi\bar{\epsilon}_i}{||y_{i,k_i^*}\!+x_i\!-\!x_j||^{\alpha}\!+\!\xi\bar{\epsilon}_i}\,dx
\right\}\!\!\right\}\\\notag
&\overset{(c)}{\approx}\mathbb{E}_{x_i}\!\left\{\exp\left\{-\lambda_c\int_{||x||>\bar{\mathcal{R}}}
\frac{\xi\bar{\epsilon}_i}{||x_i\!-\!x_j||^{\alpha}+\xi\bar{\epsilon}_i}\,dx
\right\}\right\},
\end{align}
where $(a)$ follows from taking expectation with respect to the small scale fading; $(b)$ follows from applying the pgfl of $\Phi_c$; and $(c)$ follows from using the mean value of $y_{i,k_i^*}$ to approximate the expectation with respect to $y_{i,k_i^*}$, note that the accuracy of this approximation has been verified in \cite{sys2019PCP}.

Finally, by noting that $\int_{||x||>\bar{\mathcal{R}}}=\int_{||x||>0}-\int_{||x||<\bar{\mathcal{R}}}$, the proof is complete.
\end{IEEEproof}

\begin{Remark}
{\color{black}The analytical results shown in (\ref{NOMA_outage_Pro}) can be efficiently calculated by using accurate
approximation such as the Gaussian-Chebyshev method \cite{hildebrand1987introduction} for the integrations, getting rid of exhaustive simulations for performance evaluation.}
\end{Remark}

\section{Numerical Results}
In this section, numerical results are presented to verify the analytical results and demonstrate the performance achieved by N-NOMA.
Unless stated otherwise, the parameters are set as follows: $f_c=2\times 10^9$Hz, the thermal noise power is $-170$ dBm/Hz, the transmission bandwidth is $B=10$ MHz, $\alpha=4$, $\beta_0^2=4/5$ and $\beta_1^2=1/5$, $\mathcal{R}_c=30$ m, $\mathcal{R}_{\mathcal{D}}=500$ m, $\bar{\mathcal{R}}=100$ m, $P_s=30$ dBm. Note that the target rate of the NOMA users are set as the same.
\begin{figure}[!t]
\vspace{-1em}
\setlength{\abovecaptionskip}{0em}   %
\setlength{\belowcaptionskip}{-1em}   %
\centering
\includegraphics[width=3in]{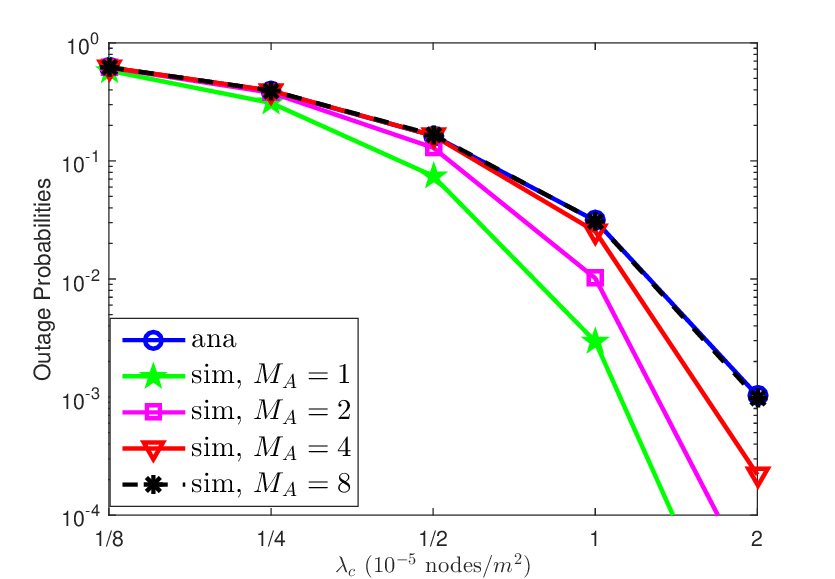}
\caption{Outage probabilities achieved by the CoMP user. $R_0=0.5$ bits per channel use (BPCU), $N=10$, $K_A=5$.}
\label{accuracy_CoMP}
\end{figure}

Fig. \ref{accuracy_CoMP} shows the outage probability achieved by the CoMP user. Note that the outage probabilities decrease with $\lambda_c$, since a larger $\lambda_c$ offers higher probability that the CoMP user can be served by more BSs. It can be observed that
the accuracy of the analytical results depends on $\lambda_c$ when $M_A$ is given. Specifically, the larger $\lambda_c$ is, the larger $M_A$ is required to be for accuracy. It is also shown that the accuracy of the analytical results can be guaranteed by a small $N$ and $K_A$.

\begin{figure}[!t]
\vspace{-0em}
\setlength{\abovecaptionskip}{0em}   %
\setlength{\belowcaptionskip}{-1em}   %
\centering
\includegraphics[width=3in]{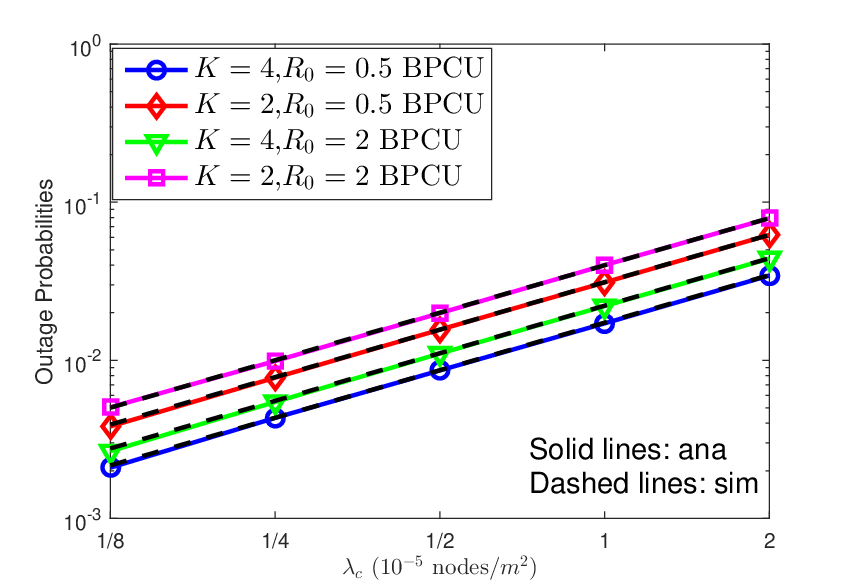}
\caption{Outage probabilities achieved by the NOMA user. $R_i=1.5$ BPCU, $N=10$.}
\label{accuracy_NOMA}
\end{figure}

Fig. \ref{accuracy_NOMA} shows the outage probability achieved by a NOMA user which is randomly chosen from $\mathcal{C}$. Simulations perfectly match analytical results which verifies the accuracy of the analysis. It is also shown that the outage probabilities achieved by NOMA users increase with $\lambda_c$, because the interferences become more severe.

\begin{figure}[!t]
\vspace{-0em}
\setlength{\abovecaptionskip}{0em}   %
\setlength{\belowcaptionskip}{-1em}   %
\centering
\includegraphics[width=3in]{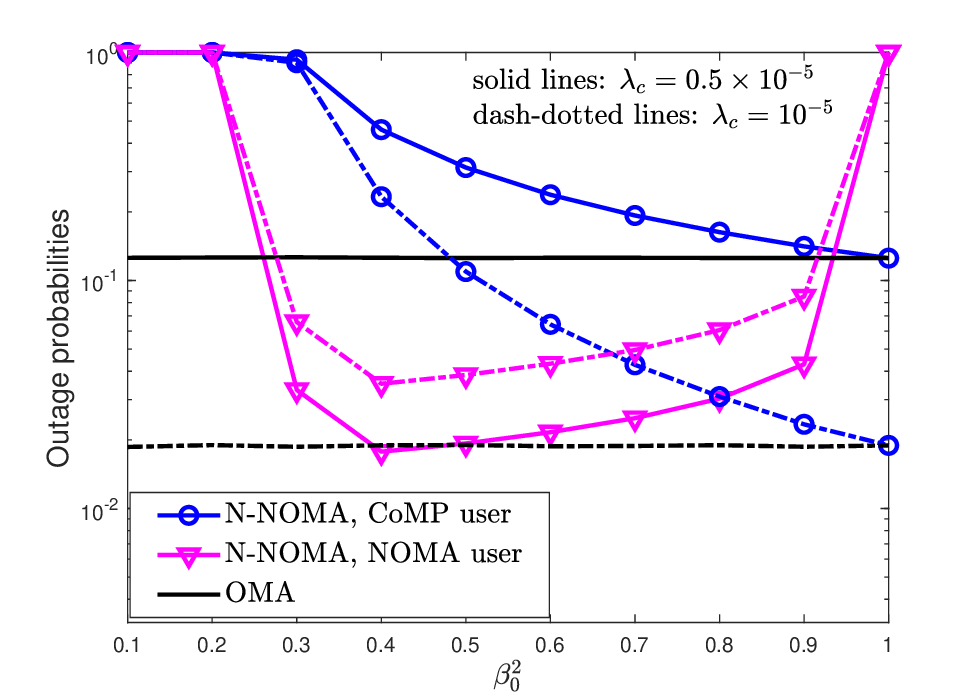}
\caption{Impact of power allocation coefficients on outage probabilities. $R_0=0.5$ BPCU, $R_i=3$ BPCU, $K=2$.}
\label{impact_beta}
\end{figure}

{\color{black}Fig. \ref{impact_beta} shows how power allocation coefficients impact the outage probabilities achieved by CoMP and NOMA users. It is shown that the CoMP user's outage probability decreases with $\beta_0$. In contrast, the outage probability achieved by NOMA user first decreases with $\beta_0$ and then increases.
And it is not hard to find from (\ref{Def_P_i}) that the turning point is $\beta_0^2=\frac{\epsilon_0+\epsilon_i\epsilon_0}{\epsilon_0+\epsilon_i\epsilon_0+\epsilon_i}$. This observation is useful for power allocation optimization.}



\begin{figure} [!t]
\vspace{-1em}
\setlength{\abovecaptionskip}{0em}   %
\setlength{\belowcaptionskip}{-1em}   %
\centering
\subfloat[Outage sum rates per BS]{\includegraphics[width=3in]{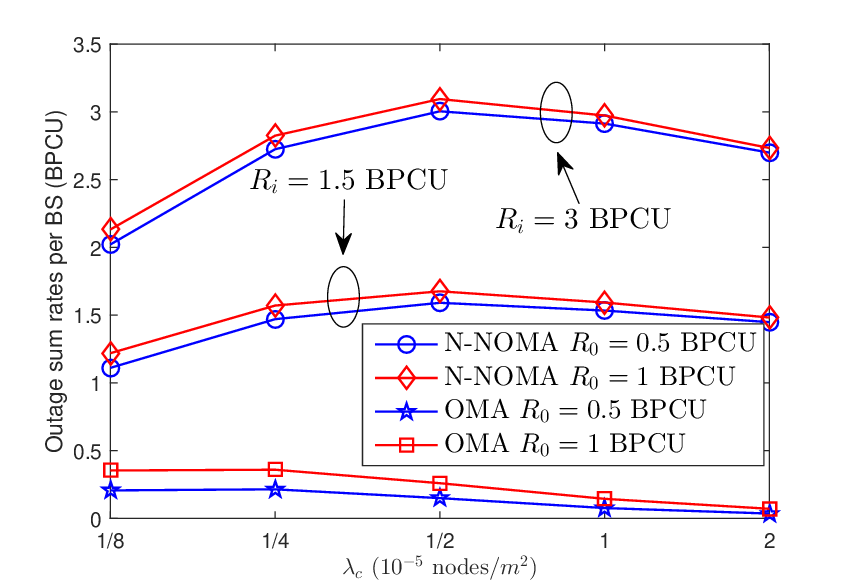}}
\\
\vspace{-1em}
\subfloat[Outage probabilities]{\includegraphics[width=3in]{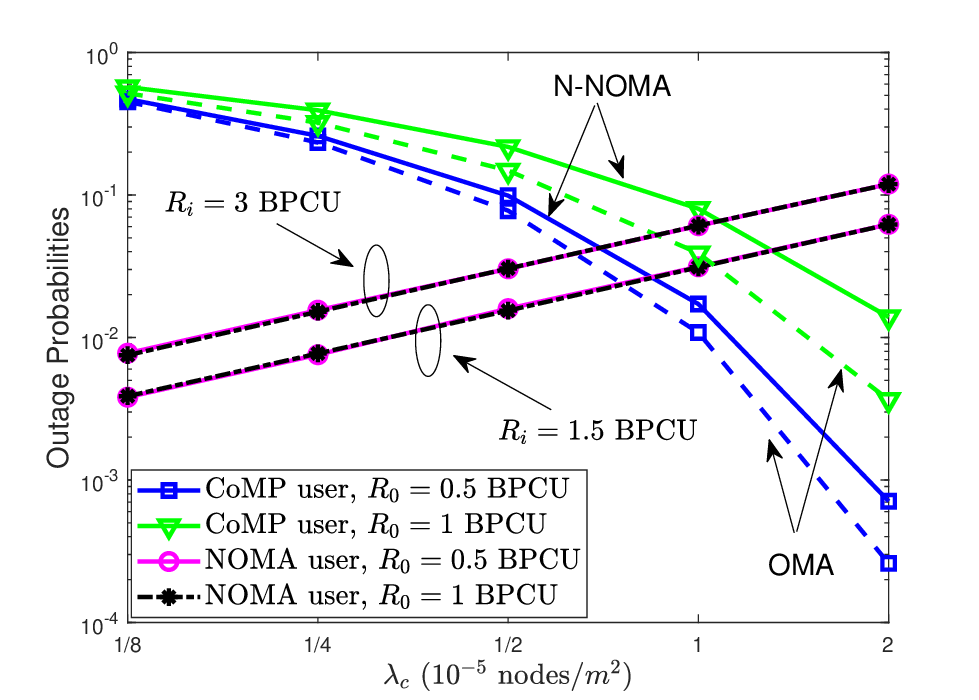}}
\caption{Comparison between N-NOMA and OMA. $K=2$.}
\label{compare}
\end{figure}

{\color{black}Fig. \ref{compare} shows the comparison between N-NOMA and OMA. Note that, in OMA, only the CoMP user is served. Fig. \ref{compare}(a) shows the outage sum rate per BS achieved by N-NOMA and OMA, and Fig. \ref{compare}(b) shows the corresponding outage probabilities
achieved by these schemes. The outage sum rate per BS is given by:
$\frac{(1-P_0^{out})R_0}{\lambda_c\pi(\mathcal{R}_D^2-\bar{\mathcal{R}}^2)}+P(M>0)(1-P_i^{out})R_i$, where
$\lambda_c\pi(\mathcal{R}_{\mathcal{D}}^2-\bar{\mathcal{R}}^2)$ denotes the average number of cooperating BSs in ring $\mathcal{C}$. Fig. \ref{compare}(a) shows that the outage sum rate per BS achieved by N-NOMA is much higher than that of OMA.
However, as shown in Fig. \ref{compare}(b), the increase of sum rate achieved by N-NOMA is at the expense of
a bit outage performance loss compared to OMA.
Thus, the condition of applicability for N-NOMA is that when the CoMP user can tolerate a bit reliability loss compared to OMA.
It also shows that the outage sum rate per BS doesn't keep increasing as $\lambda_c$ increases. Because when $\lambda_c$ is large enough, the outage probability of the CoMP user approaches zero, and the outage probabilities of NOMA users will increase with $\lambda_c$, which results in low rate.}
\section{Conclusion}
In this paper, the application of N-NOMA to a downlink CoMP system has been studied.
Stochastic geometry has been applied to model the random locations of communication nodes, based on which the outage performance of the proposed N-NOMA scheme has been evaluated. It has been shown that, by applying N-NOMA, the outage sum rate can be significantly improved compared to conventional OMA based CoMP scheme.
It is noteworthy that perfect fronthaul/backhaul capacity has been assumed in this paper, and taking the impact of limited fronthaul/backhaul capacity on performance analysis into consideration will be an interesting future extension for this paper. {\color{black}Moreover, due to the complexity of the analytical results, corresponding optimization hasn't been considered to further improve the performance. Thus, finding proper approximations to give succinct  expressions for the outage probabilities to further optimize parameters,  such as $\mathcal{R}_{\mathcal{D}}$ and $\lambda_c$, is also left as an important future work.}
\bibliographystyle{IEEEtran}
\bibliography{IEEEabrv,ref}

\begin{thebibliography}{10}
\providecommand{\url}[1]{#1}
\csname url@samestyle\endcsname
\providecommand{\newblock}{\relax}
\providecommand{\bibinfo}[2]{#2}
\providecommand{\BIBentrySTDinterwordspacing}{\spaceskip=0pt\relax}
\providecommand{\BIBentryALTinterwordstretchfactor}{4}
\providecommand{\BIBentryALTinterwordspacing}{\spaceskip=\fontdimen2\font plus
\BIBentryALTinterwordstretchfactor\fontdimen3\font minus
  \fontdimen4\font\relax}
\providecommand{\BIBforeignlanguage}[2]{{%
\expandafter\ifx\csname l@#1\endcsname\relax
\typeout{** WARNING: IEEEtran.bst: No hyphenation pattern has been}%
\typeout{** loaded for the language `#1'. Using the pattern for}%
\typeout{** the default language instead.}%
\else
\language=\csname l@#1\endcsname
\fi
#2}}
\providecommand{\BIBdecl}{\relax}
\BIBdecl

\bibitem{irmer2011coordinated}
R.~Irmer, H.~Droste, P.~Marsch, M.~Grieger, G.~Fettweis, S.~Brueck, H.-P.
  Mayer, L.~Thiele, and V.~Jungnickel, ``{Coordinated multipoint: Concepts,
  performance, and field trial results},'' \emph{{IEEE} Commun. Mag.}, vol.~49,
  no.~2, pp. 102--111, Feb. 2011.

\bibitem{maccartney2019millimeter}
G.~R. MacCartney and T.~S. Rappaport, ``{Millimeter-wave base station diversity
  for 5G coordinated multipoint (CoMP) applications},'' \emph{IEEE Trans.
  Wireless Commun.}, vol.~18, no.~7, pp. 3395--3410, Jul. 2019.

\bibitem{choi2014non}
J.~Choi, ``{Non-orthogonal multiple access in downlink coordinated two-point
  systems},'' \emph{{IEEE} Commun. Lett.}, vol.~18, no.~2, pp. 313--316, Feb.
  2014.

\bibitem{ali2018downlink}
M.~S. Ali, E.~Hossain, A.~Al-Dweik, and D.~I. Kim, ``{Downlink power allocation
  for CoMP-NOMA in multi-cell networks},'' \emph{IEEE Trans. Commun.}, vol.~66,
  no.~9, pp. 3982--3998, Sep. 2018.

\bibitem{sys2017nnomafeasibility}
Y.~Sun, Z.~Ding, X.~Dai, and G.~K. Karagiannidis, ``{A feasibility study on
  network NOMA},'' \emph{{IEEE} Trans. Commun.}, vol.~66, no.~9, pp.
  4303--4317, Sep. 2018.

\bibitem{elhattab2022joint}
M.~Elhattab, M.~A. Arfaoui, and C.~Assi, ``{Joint Clustering and Power
  Allocation in Coordinated Multipoint Assisted C-NOMA Cellular Networks},''
  \emph{IEEE Trans. Commun.}, 2022, to be published.

\bibitem{elhattab2022ris}
M.~Elhattab, M.~A. Arfaoui, C.~Assi, and A.~Ghrayeb, ``{RIS-Assisted Joint
  Transmission in a Two-Cell Downlink NOMA Cellular System},'' \emph{{IEEE} J.
  Select. Areas Commun.}, vol.~40, no.~4, pp. 1270--1286, apr 2022.

\bibitem{wang2020power}
H.~Wang, C.~Liu, Z.~Shi, Y.~Fu, and R.~Song, ``{Power minimization for two-cell
  IRS-aided NOMA systems with joint detection},'' \emph{IEEE Commun. Lett.},
  vol.~25, no.~5, pp. 1635--1639, 5 2020.

\bibitem{elhattab2020joint}
M.~Elhattab, M.~A. Arfaoui, and C.~Assi, ``{A joint CoMP C-NOMA for enhanced
  cellular system performance},'' \emph{IEEE Commun. Lett.}, vol.~24, no.~9,
  pp. 1919--1923, Sep. 2020.

\bibitem{haenggi2012stochastic}
M.~Haenggi, \emph{{Stochastic geometry for wireless networks}}.\hskip 1em plus
  0.5em minus 0.4em\relax Cambridge, U.K.: Cambridge Univ. Press, 2012.

\bibitem{sys2019PCP}
Y.~Sun, Z.~Ding, X.~Dai, and O.~A. {Dobre}, ``{On the performance of network
  NOMA in uplink CoMP systems: a stochastic geometry approach},'' \emph{{IEEE}
  Trans. Commun.}, vol.~67, no.~7, pp. 5084--5098, Jul. 2019.

\bibitem{zhang2020performance}
Y.~Zhang, J.~Mu, and J.~Xiaojun, ``Performance of multi-cell mmwave noma
  networks with base station cooperation,'' \emph{IEEE Commun. Lett.}, vol.~25,
  no.~2, pp. 442--445, feb 2021.

\bibitem{sun2021outage}
Y.~Sun, Z.~Ding, and X.~Dai, ``{On the outage performance of network NOMA
  (N-NOMA) modeled by poisson line cox point process},'' \emph{{IEEE} Trans.
  Veh. Technol.}, vol.~70, no.~8, pp. 7936--7950, 8 2021.

\bibitem{elhattab2020comp}
M.~Elhattab, M.-A. Arfaoui, and C.~Assi, ``{CoMP transmission in downlink
  NOMA-based heterogeneous cloud radio access networks},'' \emph{IEEE Trans.
  Commun.}, vol.~68, no.~12, pp. 7779--7794, Dec. 2020.

\bibitem{tanbourgi2014tractable}
R.~Tanbourgi, S.~Singh, J.~G. Andrews, and F.~K. Jondral, ``{A tractable model
  for noncoherent joint-transmission base station cooperation},'' \emph{IEEE
  Trans. Wireless Commun.}, vol.~13, no.~9, pp. 4959--4973, sep 2014.

\bibitem{sys2022QD}
Y.~Sun, Z.~Ding, X.~Dai, M.~Zhou, and Z.~Ding, ``{On the application of
  quasi-degradation to network NOMA in downlink CoMP systems},'' \emph{IEEE
  Trans. Wireless Commun.}, early access, DOI: 10.1109/TWC.2023.3284834.

\bibitem{hildebrand1987introduction}
F.~B. Hildebrand, \emph{{Introduction to numerical analysis}}.\hskip 1em plus
  0.5em minus 0.4em\relax New York, NY, USA: Dover, 1987.

\bibitem{ding2018coexistence}
Z.~Ding, P.~Fan, and H.~V. Poor, ``{On the coexistence between full-duplex and
  NOMA},'' \emph{IEEE Wireless Commun. Lett.}, vol.~7, no.~5, pp. 692--695, May
  2018.

\bibitem{sysmmwave2018}
Y.~{Sun}, Z.~{Ding}, and X.~{Dai}, ``{On the performance of downlink NOMA in
  multi-cell mmWave networks},'' \emph{IEEE Commun. Lett.}, vol.~22, no.~11,
  pp. 2366--2369, Nov. 2018.

\end{thebibliography}
\end{document}